\documentclass[a4paper,11pt]{article}
\usepackage{aaskaiid}
\usepackage[capitalise,noabbrev,nameinlink]{cleveref}
\usepackage{xspace}
\usepackage{orcidlink}
\usepackage{ulem,comment,siunitx}
\DeclareSIUnit\jansky{Jy}

\newcommand{\lcdm}{\ensuremath{\Lambda}\rm CDM\xspace}

\newcommand{\hi}{\textrm{H\textsc{i}}\xspace}

\title{Overview: Cosmology with the SKAO}
\ShortTitle{Cosmology Overview}

\ShortName{Cosmology SWG Chairs} % shortened name list for header 
\author[1,2]{Marta Spinelli\orcidlink{0000-0003-0148-3254}}
\author[3]{Ian Harrison\orcidlink{0000-0002-4437-0770}}
\author[4]{Richard A.\ Battye\orcidlink{0000-0002-4269-515X}}
\author[5,6,7,2]{Stefano Camera\orcidlink{0000-0003-3399-3574}}
\author[8,9]{Xulei Chen\orcidlink{0000-0001-6475-8863}}
\author[2,10]{Roy Maartens\orcidlink{0000-0001-9050-5894}}
\author[2,11]{Mario G.\ Santos\orcidlink{0000-0003-3892-3073}}
\author[4]{Laura Wolz\orcidlink{0000-0003-3334-3037}}

\affiliation[1]{Observatoire de la Côte d’Azur, Laboratoire Lagrange, Bd de l’Observatoire, 06304 Nice, France}
\affiliation[2]{Department of Physics \& Astronomy, University of the Western Cape, 7535 Cape Town, South Africa}

\affiliation[3]{School of Physics and Astronomy, Cardiff University, Cardiff CF24 3AA, UK}

\affiliation[4]{Jodrell Bank Centre for Astrophysics, Department of Physics and Astronomy, University of Manchester, Manchester M13 9PL, U.K}

\affiliation[5]{Dipartimento di Fisica, Universit\`a degli Studi di Torino, Via P.\ Giuria 1, 10125 Torino, Italy}
\affiliation[6]{INFN -- Istituto Nazionale di Fisica Nucleare, Sezione di Torino, Via P.\ Giuria 1, 10125 Torino, Italy}
\affiliation[7]{INAF -- Istituto Nazionale di Astrofisica, Osservatorio Astrofisico di Torino, Strada Osservatorio 20, 10025 Pino Torinese, Italy}

\affiliation[8]{State Key Laboratory of Radio Astronomy and Technology, National Astronomical Observatories, Chinese Academy of Sciences, 20A Datun Road, Chaoyang District, Beijing 100101, China}
\affiliation[9]{School of Astronomy and Space Science, University of Chinese Academy of Sciences, Huairou District, Beijing 101408, China}

\affiliation[10]{National Institute for Theoretical \& Computational Sciences, Cape Town 7535, South Africa}

\affiliation[11]{South African Radio Astronomy Observatory (SARAO), Cape Town, 7700, South Africa}

\abstract{The SKA telescopes will revolutionise our ability to do cosmology at radio wavelengths, via both their own data and in synergy with other wavelengths. SKAO will be the first instrument able to conduct large-scale cosmological surveys as done in the last decades in the optical and near-infrared. This complementarity will be vital as cosmology hits the limit of systematic uncertainties. Radio cosmology surveys will have radically different systematics, allowing data combinations across surveys to calibrate systematics and increase overall constraining power. Neutral hydrogen (\hi) intensity mapping surveys are now reaching maturity, as demonstrated by the progress made by the MeerKLASS survey with MeerKAT. Along with continuum galaxy surveys, they  will provide detailed maps of the Universe covering large fractions of the sky, allowing us to answer questions about fundamental physics which can only be measured on the largest scales. In combination with weak lensing and \hi galaxy probes, \hi intensity maps will also measure the distributions of matter and velocities to give precisions tests of the \lcdm model, including its foundational assumptions of isotropy and homogeneity.  In combination with gravitational wave observations and fast radio bursts, they will also help us measure the expansion history and baryon content of the Universe. Here we provide an overview of the achievements of precursor surveys and the progress towards SKA cosmology,  starting with AA* and reaching full maturity with AA4 telescopes.}

\begin{document}
\maketitle

\section{Introduction}
\label{sec:introduction}

The prospect of using the Square Kilometre Array Observatory (SKAO) telescopes to do wide area cosmology surveys competitive with those at other wavebands remains a tantalising one. Cosmology seeks to infer the underlying fundamental physics that shapes the origin and evolution of the Universe as a whole. SKAO surveys will provide unprecedented access to large volumes in the Universe, allowing us to uniquely probe the largest scales, as well as providing ways to break through the limiting ceiling of systematic uncertainties affecting all cosmological experiments.

Since the previous SKA Science Book \citep[in which the cosmology overview is given by][]{Maartens:2015mra}, evolution of the ``Standard'' or ``Concordance'' cosmological model has been subtle but significant. The validity of the \lcdm model -- describing a nearly homogeneous and isotropic expanding space-time governed by Einstein's General Relativity with baryonic matter, dark matter (DM) and dark energy (DE), evolving from initial perturbations generated by inflation --  still stands.

However, multiple questions remain about the true physical nature of DM and DE, and whether GR holds on all scales. At the same time, the most recent cosmological surveys have collected data that the standard theory cannot fully describe. This puzzling picture goes from the $H_0$ tension \citep[e.g.][]{Knox:2019rjx} to the tentative evidence for evolution of DE \citep{DESI:2025zgx}, showing that we cannot currently fully describe why the expansion history of the Universe proceeds as it does. Moreover, the measurements of the spectrum of primordial fluctuations have begun to severely impinge on previously favoured models of inflation \citep{AtacamaCosmologyTelescope:2025nti,Ferreira2026}, and upper limits on the total mass of neutrinos have become uncomfortably close to the lower limits provided by particle physics \citep{SPTCamphuis2026,AtacamaCosmologyTelescope:2025nti}. Cosmological results are also increasingly revealing the importance of astrophysical processes in their interpretation. As an example, the growth of structures on quasi-non-linear scales is still incompletely understood, with the interplay of active galactic nuclei (AGN) feedback, exotic matter components, and modified gravity unclear -- even if this gap has narrowed as measurements of $S_8$ from different redshifts and length scales have progressed \citep[see][for an up to date review]{Pantos:2026koc}.

SKAO surveys can provide vital input to the above challenges, across a range of different types of observation. Where SKAO surveys use radio emissions as a tracer of cosmological physics, they can provide unique information by probing the largest observational volumes, by helping beat systematic uncertainties, and by expanding the potential discovery space.

The practical realisation of such surveys has been continually advanced in the past ten years by SKAO pathfinder and precursor experiments. \hi intensity mapping has been demonstrated on ever larger scales across multiple different instruments \citep[][]{Cunnington01.2026.SKA,Elahi01.2026.SKA}, continuum galaxy surveys have provided ever wider and deeper views of the Universe \citep{Asorey01.2026.SKA}, and methods to detect the large-scale gravitational lensing of galaxy shapes have been developed and tested on real data \citep{Harrison02.2026.SKA}. \hi galaxy surveys have begun to probe beyond the local volume, allowing measurements which track the expansion history (via the Tully-Fisher relationship) and probe the growth of structure through the velocities of matter in the Universe \citep{Nasirudin01.2026.SKA,Mayor01.2026.SKA}.

In addition, entirely new and unique ways of probing cosmology with radio observations have  emerged. Fast radio bursts (FRBs) \citep{Caleb01.2026.SKA} probe large-scale distributions of ionised media, allowing them to constrain both structure growth and background expansion, and the first observations of gravitational waves have led to a surge in interest in using them along with the radio to also probe these observables \citep{Baker01.2026.SKA}.

The AA* configuration will be already capable of delivering novel cosmological results, while the power of wide-area scanning at high sensitivity and high angular resolution with the full AA4 configuration will allow the SKAO to become a fully-fledged instrument for cosmology, alongside leading facilities in other wavebands such as \textit{Euclid}, Rubin, Simons Observatory, Roman and Cosmic Explorer. Cosmology is therefore a key driver on the path to realising the full potential of the SKAO, providing both precise and timely technical feedback during its initial phases and a vibrant theoretical framework to support the continued growth and evolution of the Observatory.

% \section{The SKA Cosmology SWG}
\section{Observational Themes in Radio Cosmology}
\label{sec:theswg}
Here we delineate the different types of observational methods and the corresponding science themes relevant to SKAO Cosmology. These themes have been actively and continually evolving throughout the development of the telescopes, driven by ambitious overarching cosmology science goals, fundamental physical questions which can be answered by the SKAO and which we discuss in \cref{sec:goals}. These have provided novel observational strategies and stringent instrumental requirements throughout the history of the Observatory.

The themes naturally map to different focus areas for different members of the science community, spanning observational astronomers, survey scientists, data analysts, instrumentalists, and theoretical cosmologists, and reflect the organisation of the SKA Cosmology Science Working Group (SWG). The radio cosmology community contributed extensively to the landmark \textit{Advancing Astrophysics with the SKA Science Book} published in 2014, with several dedicated chapters addressing key cosmological probes and methodologies \citep[with an overview provided by][]{Maartens:2015mra}. Building upon this foundational effort, the SWG later coordinated the production of the “Cosmology with Phase 1 of the Square Kilometre Array – Red Book” \citep{RedBook}, a major reference document that established the technical specifications and survey strategies for most cosmological forecasts, in particular with SKA-Mid. The Red Book represented a milestone for the community, consolidating years of collaborative work into a coherent roadmap for precision cosmology with the SKAO.

% The SKA Cosmology Science Working Group (SWG) has been an active and scientifically broad community within the SKAO ecosystem since the earliest conceptual phases of the Observatory. Established to develop and coordinate the cosmological science case, the SWG has progressively grown into a vibrant international collaboration comprising more than 300 researchers from institutes across the world. The group brings together experts spanning a remarkably wide range of expertise, including observational astronomers, survey scientists, data analysts, instrumentalists, and theoretical cosmologists. This heterogeneous composition has been one of the major strengths of the community, enabling a continuous dialogue between theoretical ambitions, observational strategies, and instrumental requirements throughout the development of the Observatory. Over the years, the SWG has continuously evolved in response to both theoretical developments and the increasing maturity of the Observatory.

\begin{figure}[h]
    \centering
	\includegraphics[width=0.9\textwidth]{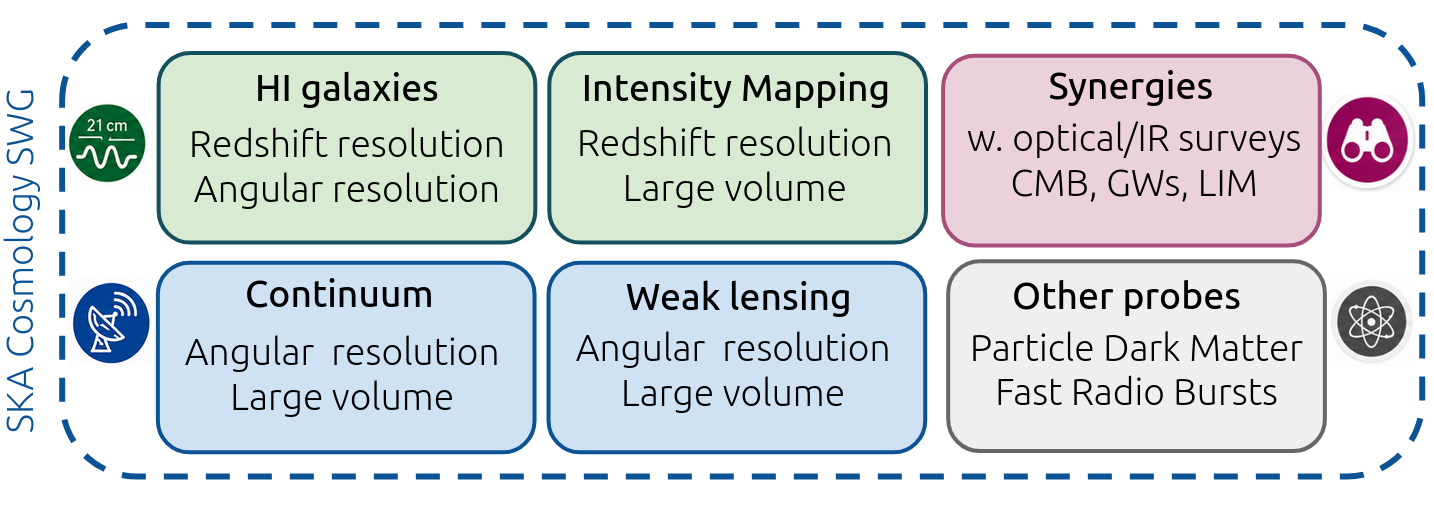}
    \caption{A map of the established thematic areas in radio cosmology, plus a new growing community proposing other probes. The schematic is colour coded according to the main probe considered: the \hi in green, radio continuum in blue, synergies in magenta. For each area, a very synthetic description is presented. See the text for more details.}
    \label{fig:swg}
\end{figure}

The community now covers a broad range of scientific topics, from dark energy and modified gravity to primordial non-Gaussianity, ultra-large scales, neutrino physics, intensity mapping, weak lensing, radio continuum cosmology, and synergies with other wavelengths. The thematic structure of SKAO Cosmology, discussed below and illustrated in \autoref{fig:swg}, provides the framework for collaboration, scientific planning, and community-driven development within the Cosmology SWG. Here we provide short summaries of the science within each thematic area and the chapters in the book which are most closely aligned with their work.

\subsubsection*{\hi Galaxies}

One of the key science themes in radio cosmology is the use of \hi\ spectral line observations of individual galaxies across wide fields, with the long-term goal, explored in theoretical studies, of a fully spectroscopic ``billion galaxy survey'' using a future version of the SKAO.

The detection of the 21\,cm emission line of neutral hydrogen has an exquisite intrinsic spectral resolution that allows extremely accurate redshift measurements. HI galaxy surveys thus provide a direct spectroscopic mapping of large-scale structure, enabling precise measurements of the growth of cosmic structure, baryon acoustic oscillations (BAO), and redshift-space distortions. In addition, these \hi observations probe the cold gas content of galaxies, directly linking cosmology to galaxy evolution and the baryon cycle. In this book:
\begin{itemize}
    \item \citet{Nasirudin01.2026.SKA} forecasts how well SKAO surveys will constrain cosmological parameters using the BAO scale, redshift-space distortions, and the matter power spectrum turnover scale.
    \item \citet{Mayor01.2026.SKA} shows the detectability of the Tully-Fisher effect and discusses its use in measuring the expansion history of the Universe and using peculiar velocities to map large scale structure growth.
    \item \citet{Ronconi01.2026.SKA} discusses simulations of the \hi content of the Universe and the interplay between simulations and observations.
\end{itemize}

\subsubsection*{\hi Intensity Mapping}

Complementing galaxy redshift surveys, \hi\ intensity mapping (IM) probes the large-scale distribution of neutral hydrogen without detecting individual galaxies, requiring advances in both theoretical modelling and data analysis methodologies.

\hi IM is of particular interest for the cosmological community as it allows us to probe the large-scale structure across unprecedented volumes. Rather than resolving individual galaxies, \hi IM measures the collective 21\,cm emission from unresolved neutral hydrogen, enabling efficient three-dimensional mapping of matter fluctuations over wide redshift ranges. As for \hi galaxy surveys, IM provides access to key observables such as BAO and redshift-space distortions, but at higher redshift. Moreover, large volumes allow access to the largest scales where it is possible to test primordial non-Gaussianity and ultra-large-scale relativistic effects. With SKA-Mid, \hi IM is expected to probe the large-scale neutral hydrogen distribution up to redshifts $z\sim 3$, while SKA-Low, although primarily targeting the Epoch of Reionization, naturally covers the intermediate range $3\lesssim z\lesssim6$, which remains highly relevant for studies of structure growth and the evolution of neutral hydrogen after reionization. Together, SKA-Mid and SKA-Low offer the prospect of a view of cosmic structure from the local Universe deep into the reionization era, further motivating ongoing theoretical development and simulation efforts within the IM community.
Over the past decade, \hi intensity mapping has evolved into a highly active field, driven by a growing effort from pathfinder and precursor.
In particular, the MeerKLASS survey \citep{santos2015} has established single-dish intensity mapping with MeerKAT, demonstrating the feasibility and the interest of a similar but deeper survey with SKA-Mid.

In this book:
\begin{itemize}
    \item \citet{Wolz01.2026.SKA} reviews the full science case for \hi IM in measuring cosmological information.
    \item \citet{Cunnington01.2026.SKA} summarises results from the MeerKLASS survey, highlighting its success in building single-dish IM calibration pipeline for MeerKAT and in achieving the first large-scale detections of \hi. 
    \item \citet{Elahi01.2026.SKA} reports results and lessons learned in \hi IM from other SKA pathfinders (CHIME, FAST, GBT, GMRT, HIRAX, Parkes, Tianlai).
    \item \citet{Spinelli01.2026.SKA} sets out current methods for dealing with the strong foregrounds and observational systematics that must be  tightly controlled for \hi IM.
    \item \citet{Mazumder01.2026.SKA} forecasts the cosmology which can be achieved by performing \hi IM in interferometric observing modes, and reviews current results from MeerKAT.
    \item \citet{Majumdar01.2026.SKA} shows the cosmology available by considering higher-order non-Gaussian statistics of \hi IM, discussing different ways of measuring this extra information.
\end{itemize}

In addition, we highlight \citet{Chatterjee01.2026.SKA} as a highly important methodological contribution, which discusses how on-the-fly observing modes can be used to simultaneously generate continuum imaging while performing fast constant elevation scanning for single-dish \hi intensity mapping.

\subsubsection*{Continuum Galaxies}

Radio continuum surveys of individual galaxies provide a powerful probe of cosmology through their number counts and clustering over wide areas of the sky.
As well as providing measurements of galaxy bias and tracing large-scale structure at low redshift, a significant area of activity is in measuring the cosmic dipole and using the observations in conjunction with the Continuum Galaxies SWG to infer the galaxy and star formation history of the Universe. In this book:
\begin{itemize}
    \item \citet{Asorey01.2026.SKA} showcases the clustering of continuum galaxies and how well SKA-Mid will be able to use it to measure bias and redshift properties of the sources, and measure cosmology power spectra.
    \item \citet{Bertacca01.2026.SKA} shows forecasts for the measurement of the large-scale isotropy and homogeneity of the Universe with SKA sources, both in number counts and kinematics, testing one of the fundamental assumptions of cosmology.
\end{itemize}

As already mentioned, \citet{Chatterjee01.2026.SKA} present on-the-fly mapping techniques that allows commensal fast survey speed for deep and large area continuum images for MeerKAT. The lessons learned from this technique will be valuable for the upcoming SKA-Mid, where better resolution
and sensitivity are expected. 

In addition, \citet{Harrison01.2026.SKA} shows how SKA surveys may be combined together in so-call $N\times2$pt surveys which use galaxy clustering, weak lensing and \hi probes together to calibrate systematics and increase cosmological constraining power.

\subsubsection*{Weak Lensing}
Wide-field observations of resolved radio continuum galaxies enable weak gravitational lensing measurements over large areas of the sky. These measurements trace the gravitational lensing potential, an unbiased tracer of the matter distribution and hence an excellent probe of the physics of structure growth. 

In this book:
\begin{itemize}
    \item \citet{Harrison02.2026.SKA} shows the utility for cosmology of SKAO weak lensing cosmic shear surveys.
    \item \citet{Tripathi01.2026.SKA} compares source separation and radio shape measurement techniques, including the most recent development based on Deep Learning.
    \item \citet{Pandey-Pommier02.2026.SKA} discusses how SKAO will transform cluster lensing studies enabling high-fidelity reconstruction of cluster mass distributions and revealing a previously inaccessible population of faint galaxies.
\end{itemize}

\subsubsection*{Synergies}

Multi-probe cosmology exploits the complementarity between the observational approaches described above, as well as with external datasets, to deliver tighter constraints on cosmological models.
In particular, the different clustering bias and observational systematics make cross-correlations with optical surveys especially powerful for calibrating galaxy bias, improving measurements of structure growth, and implementing multi-tracer analyses on ultra-large scales. The complementarity with surveys such as \textit{Euclid} and LSST is a recurring theme of the SKA cosmology programme, since radio and optical observations provide independent probes of the same underlying matter distribution with substantially different systematics.

In this book:
\begin{itemize}
    \item \citet{Camera01.2026.SKA} discusses the many ways in which SKA observations, in combination with optical surveys, will be able to test the theory of gravity on large scales in the Universe.
    \item \citet{Fonseca01.2026.SKA} shows the ability of SKA surveys, alone and in combination with optical surveys, to shed light on the very high energy physics of the earliest times in the Universe, including primordial non-Gaussianity.
    \item \citet{Baker01.2026.SKA} discusses the synergy with gravitational wave observations in constraining expansion history and structure formation.
    \item \citet{Sarkar01.2026.SKA} demonstrates how the calibration of systematic effects in \hi IM can be mitigated by cross-correlation with IM of lines at other wavelengths, disentangling astrophysical and cosmological parameters.
\end{itemize}

\begin{figure}[t]
    \centering
	\includegraphics[width=\textwidth]{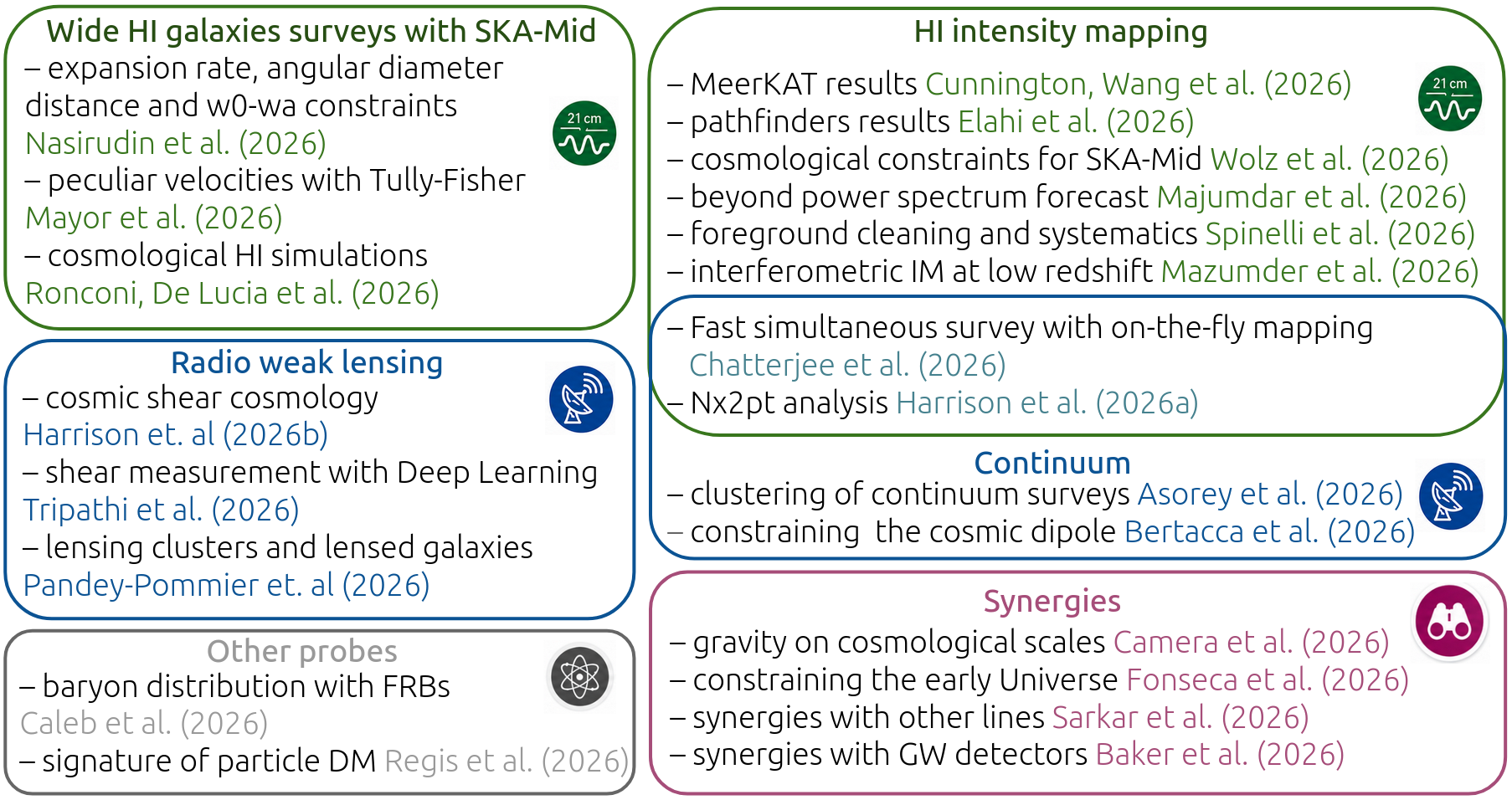}
    \caption{Schematic grouping of chapters contributed by the different thematic areas within the SKA cosmology SWG, summarising the information in \cref{sec:theswg}.}
    \label{fig:swg_chapters}
\end{figure}

\subsubsection*{Other probes}
The large number of open questions in cosmology, together with their fundamental nature, motivates the community to explore every possible new observational window. The projected sensitivities of the SKA telescopes in the AA4 baseline design have prompted the search for dark matter candidates through radio observations. In particular, signatures of WIMPs, axion-like particles, and dark photons can be investigated through a combination of spectral-line, continuum, and transient observations.

At the same time, precision cosmology requires a comprehensive understanding of the role played by astrophysical processes in cosmic evolution. Studies of the matter distribution, especially on intermediate and small scales, cannot be carried out without a detailed characterisation of baryonic physics. Once again, the broad range of observations enabled by the SKA provides a unique opportunity to bridge cosmology and astrophysics, including the study of transient phenomena such as FRBs.

In this book:
\begin{itemize}
    \item \citet{Regis01.2026.SKA} details how continuum, spectroscopic and polarisation measurements from SKA telescopes can all be used to potentially detect emission signatures from different models of particle DM.
    \item \citet{Caleb01.2026.SKA} discuss how FRBs are emerging as a highly significant probe of the baryon distribution, also improving our knowledge of DM. Note that this chapter completes an FRB trilogy. The interested reader can find the presentation of the other FRB chapters in the Transients overview: \citet{Miller-Jones01.2026.SKA}.
\end{itemize}

In \autoref{fig:swg_chapters}, we provide a map of the chapters related to cosmology that are presented in this new version of the SKA Science Book, arranged with respect to the themes just discussed and colour-coded according to \autoref{fig:swg}.

\subsection{Typical Assumed Surveys}
\label{sec:surveys}
Across the chapters, the assumed cosmological surveys can be divided into a few broad classes, summarized below. While these survey specifications correspond to the full AA4 capabilities (following the previous SKA Cosmology Red Book \citep{RedBook}), the individual chapters also discuss the scientific potential of the intermediate AA* stages.

\begin{itemize}
    \item \textbf{Mid Band 2 Medium-deep Survey:} Expected to cover around $5\,000\,\deg^2$ to a continuum depth of $\sim\qty{3}{\micro\jansky}/\mathrm{beam}$ as a series of mosaicked pointings, providing both continuum and \hi line galaxies. Additionally, many \hi IM results assume a survey of the same area and total integration time (10\,000$\,$hours) from constant elevation scanning. Both types of \hi line observations would cover the redshift range $0 < z < 0.4$. Most surveys expect this area to overlap the DES and \textit{Euclid} regions in the Southern sky, and also cover as much as possible the part of the DESI survey accessible by SKA-Mid, in order to maximise synergies with other wavebands. 
    \item \textbf{Mid Band 1 Wide Survey:} This wider survey would cover $\sim20\,000\,\deg^2$ (most of the Southern sky) over $10\,000$ hours. Again, both continuum galaxy surveys and \hi IM surveys are variously assumed, with \hi line observations covering $0.35 < z < 3$, a highly valuable redshift range for the broader community due to the lack of redshift coverage for mainstream optical cosmology surveys.
    There is much interest in allowing both types of survey to be conducted at the same time, using the on-the-fly mapping techniques pioneered in cosmology surveys with precursor telescopes \citep{Chatterjee01.2026.SKA}.
\end{itemize}

\autoref{tab:aastar} provides an indicative overview of the expected scientific reach of the different SKAO configurations for the cosmology science cases discussed in this review, illustrating the progression from the intermediate AA* stages to the full AA4 capabilities.

\begin{table}
	\centering	
	\caption{Indicative table of results which will be possible in different science cases with different SKAO configurations. Green check marks {\color{green}\checkmark} indicate that observations will be possible to their full extent. Orange check marks  {\color{orange}\checkmark} indicate that measurements will be possible with less constraining power, but fundamental to understand and control systematics (as discussed in \cref{sec:conclusions}). Dashes represent measurements which will not be possible.}
	\vspace{-0.2cm}
	\begin{tabular}{lccccc}
            \hline
            Physics & Probe & Reference & AA4 & AA* \\
		\hline
            \lcdm cosmology & WL & \citet{Harrison02.2026.SKA} & {\color{green}\checkmark} & -- \\
            \lcdm cosmology & \hi IM & \citet{Wolz01.2026.SKA} & {\color{green}\checkmark} & {\color{orange}\checkmark} \\
            \lcdm cosmology & $N\times2$pt & \citet{Harrison01.2026.SKA} & {\color{green}\checkmark} & -- \\
            Hubble constant, structure growth & \hi gals & \citet{Mayor01.2026.SKA} & {\color{green}\checkmark} & {\color{orange}\checkmark} \\
            Evolving Dark Energy & \hi gals & \citet{Nasirudin01.2026.SKA} & {\color{green}\checkmark} & {\color{orange}\checkmark} \\
            Evolving Dark Energy & \hi IM, opt.\ gal & \citet{Camera01.2026.SKA} & {\color{green}\checkmark} & {\color{orange}\checkmark} \\
            Evolving Dark Energy & $N\times2$pt & \citet{Harrison01.2026.SKA} & {\color{green}\checkmark} & -- \\
            Primordial non-Gaussianity & \hi IM, opt.\ gal & \citet{Fonseca01.2026.SKA} & {\color{green}\checkmark} & {\color{orange}\checkmark} \\
             Relativistic effects & $N\times M$pt & \citet{Camera01.2026.SKA} & {\color{green}\checkmark} & -- \\
            Cosmic dipole & cont.\ gals & \citet{Bertacca01.2026.SKA} & {\color{green}\checkmark} & -- \\
            Baryon feedback & FRBs & \citet{Caleb01.2026.SKA} & {\color{green}\checkmark} & {\color{orange}\checkmark} \\
            Particle Dark Matter & multiple & \citet{Regis01.2026.SKA} & {\color{green}\checkmark} & {\color{orange}\checkmark} \\
            \hline
		\hline
	\end{tabular}
	\label{tab:aastar}
\end{table}

\section{Key Science Goals}
\label{sec:goals}
The surveys and individual science cases described above exist typically to address the broad science questions posed in modern cosmology. As discussed in \cref{sec:introduction}, the SKA telescopes will be critical in solving modern cosmological problems due to their unique ability to probe large volumes, provide independent constraints which mitigate systematics and probe unknown discovery spaces. Here we highlight these thematic areas and the scientific questions we seek to answer.

\subsection{Probing the Largest Volumes}
\begin{figure}[h]
    \centering
	\includegraphics[width=0.8\textwidth]{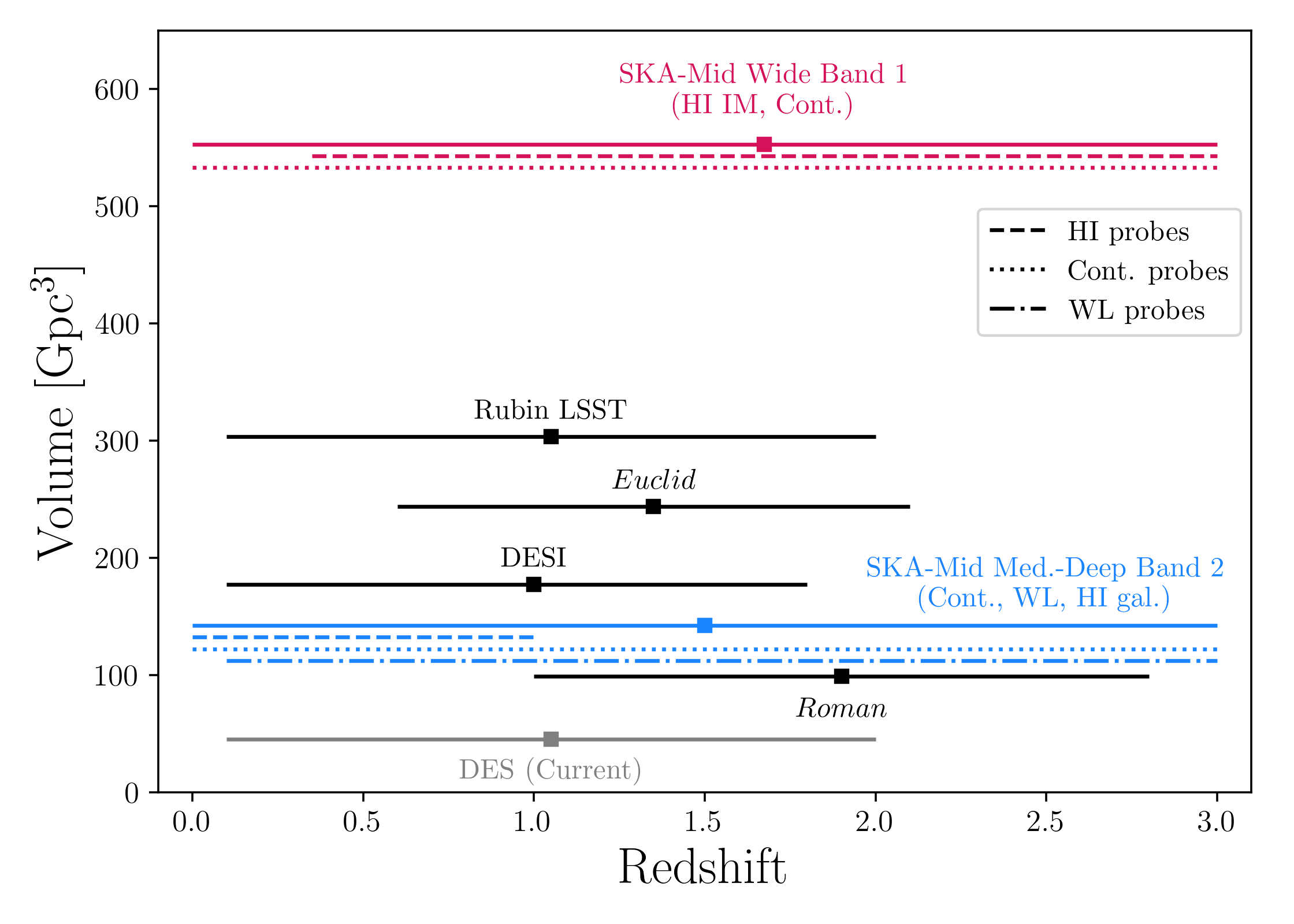}
    \caption{Indicative survey volumes for a selection of current and near-future cosmology surveys, including the two SKA-Mid surveys described in \cref{sec:surveys} assuming the AA4 configuration. For the SKA surveys we indicate which parts of the redshift range are covered by continuum galaxy, weak lensing and \hi (Intensity Mapping and Galaxy) probes. The figure is an updated version of the one in \citet{Maartens:2015mra}.}
    \label{fig:multiprobe]}
\end{figure}
As shown in \cref{fig:multiprobe]}, SKAO will have an unmatched ability to probe the largest volumes in the Universe, greater than comparable cosmology experiments in other wavebands. This will enable SKA surveys to uniquely answer questions at the forefront of cosmology.

\citet{Camera01.2026.SKA} shows how we will be able to tackle the question: \textbf{\textit{Is General Relativity correct on the largest scales?}} General Relativistic contributions to the observable density contrast of astronomical tracers are most significant on large scales, exactly the same scales on which the irreducible cosmic variance from limited numbers of independent directions on the sky sets a floor on measurement precision. SKA surveys can circumvent this through `multi-tracer' observations on large areas. By combining \hi galaxy, \hi intensity mapping and continuum galaxy observations in surveys of above $5\,000\,\deg^2$ these effects will be detectable, providing a crucial null test of General Relativity in cosmology. The phenomenological consequences of the plethora of modified gravity theories will also be observable: all three types of survey will also provide measurements of Effective Field Theory of DE and modified gravity parameters, the $\mu$ (modified Poisson) and $\eta$ (modified lensing) parameters, and the growth index $\gamma$.

Secondly, \cite{Fonseca01.2026.SKA} shows how SKAO will be able to address the question \textbf{\textit{How are structures in the Universe seeded at the earliest times?}} The current cosmological paradigm is that a period of scalar field-driven accelerated expansion, Inflation, stretched quantum mechanical density fluctuations to large scales in the initial period of the Big Bang. A key signature of exactly what kind of scalar field also lies in the distribution of astronomical objects on the very large angular scales which will be observable with SKAO. Through the use of multi-tracer analyses between SKAO \hi IM and an LSST-like photometric galaxy survey, measurements of local-type primordial non-Gaussianity $\sigma(f^{\rm local}_{\rm NL}) < 1$ will be possible, beating current limits and constraining the space of possible particle physics models which could have driven the Inflation era.

Finally, SKAO will allow us to test the basic cosmological principle: \textbf{\textit{Is the Universe statistically isotropic and homogenous?}} By measuring the dipole anisotropy of source positions and kinematics with \hi and continuum surveys, \citet{Bertacca01.2026.SKA} show how SKAO will be able to resolve current disagreements about tentative departures from the expected behaviour in current surveys.

\subsection{Beating the Systematics Limit}
Cosmology is a field where progress is increasingly limited by systematic, not statistical, uncertainties. Even in areas where the SKA telescopes will not be competitive in terms of their intrinsic signal-to-noise, they will be powerful tools to provide data-driven ways to calibrate systematics in other experiments. Without SKAO these systematics will limit our ability to learn about the fundamental nature of the Universe.

Modern astrophysics, cosmology and particle physics are all fundamentally linked in pursuit of the question: \textbf{\textit{Is dark matter truly particle in nature?}} As described in \citet{Regis01.2026.SKA}, SKA will be able to provide a unique and broad-ranging perspective on this question, being sensitive to particle dark matter candidates: WIMPs and axion-like particles. Targeted observations, time-domain observations, and high spatial and spectral resolution all contribute towards the ability of the SKA to improve constraints in multiple search directions.

Several SKAO observables will be able to help us answer the question: \textbf{\textit{How do baryonic processes affect cosmological measurements?}} Baryonic feedback -- energy injections from AGN -- can redistribute matter on scales relevant to cosmology, creating an astrophysical systematic which can bias or dilute estimates of crucial cosmological parameters related to the abundance and growth of large-scale structures (such as the $S_8$ parameter discussed in \cref{sec:introduction}). Understanding these effects has recently been identified as a crucial limiting factor on all next-generation cosmological surveys, and discerning a clear picture has proved difficult across current observations and hydrodynamical simulations. These astrophysical effects can also be highly degenerate with the effect of exotic forms of matter and energy which can explain DM and DE. SKAO will be able to probe baryonic feedback directly through the mapping of baryon distributions with FRB dispersion measures \citep{Caleb01.2026.SKA} and understand the interplay between baryons and the total matter distribution, using weak gravitational lensing  \citep{Harrison02.2026.SKA}.

The FRB probe is of course unique to radio telescopes and SKAO measurements will be capable of improving measurements of affected cosmological parameters by factors 2--5 compared to next-generation optical surveys alone. The greatest gains are made from $15\,000\,\deg^2$ surveys with the SKA-Low telescope, but similar gains may be had from a survey using Band 2 of SKA-Mid.

Weak lensing in the radio, utilising $5\,000\,\deg^2$ high resolution continuum surveys in Mid Band 2, can also provide degeneracy-breaking information on limiting systematics. As well as allowing different parts of the volume to be probed (higher redshifts and large scales), thus differentiating between theories which act differently throughout the history of the Universe, radio weak lensing can be used in cross-correlation with optical lensing to self-calibrate redshift and shear measurement nuisance parameters. Further gains can be made by using kinematic and polarisation information to measure the intrinsic alignment of galaxies which can otherwise mimic the lensing signal. This allows SKA lensing surveys to provide useful cosmological information even where they are not otherwise competitive. By measuring lensing of individual clusters \citep{Pandey-Pommier02.2026.SKA}, SKA will also be able to test baryon models in their small-scale neighbourhoods.

Similarly, SKAO surveys can help answer the question: \textbf{\textit{Why do different methods of measuring the cosmic expansion history disagree?}} The high spectroscopic precision of \hi surveys will be capable of mapping the expansion history with both the BAO standard ruler \citep{Nasirudin01.2026.SKA} and the Tully-Fisher relation \citep{Mayor01.2026.SKA}. By providing electromagnetic counterpart observations for gravitational wave standard sirens, \hi intensity mapping can potentially provide both an extra low-redshift measurement of $H_0$ to arbitrate the `Hubble tension' and a measurement of the evolution of DE at higher redshifts \citep{Baker01.2026.SKA}.

\subsection{Expanding Discovery Space}
Entirely new windows on the Universe will be enabled by SKA cosmology surveys. Wide-area surveys required for cosmology are also sources of legacy data which can be factories for new discoveries. In precursor and pathfinder cosmology-focused surveys we have already seen the discovery of new objects such as odd radio circles \citep[ORCs][]{2022MNRAS.513.1300N} and reionisation-limited \hi clouds \citep[RELHICS][]{Anand:2025czy}. By pushing into a new frontier of surveyed volume at radio wavelengths, more will surely appear.

\section{Conclusions}
\label{sec:conclusions}
The SKA Observatory will be revolutionary for cosmology: for the first time radio telescopes will be capable of large-volume surveys. When fully realised with the AA4 configuration, both continuum and \hi spectral line surveys will probe the true nature of the Universe, on the largest scales and in the greatest detail. They will have significant legacy value through comprehensive observations of the southern sky and through complementary combinations with surveys at other wavelengths.

Cosmology surveys will push the boundaries of the observation capabilities of radio telescopes. They have stringent systematics limits which require exquisite control in telescope operations and data analysis pipelines, from calibration to high-fidelity imaging.
Achieving the cosmology science goals will generate positive feedback with the SKAO and the broader science community and will help in achieving the highest science return from the telescopes.

Some of the cosmology science cases for SKAO rely on observations which will not be possible without the full deployment of AA4, as illustrated in \cref{tab:aastar}. For example, science results involving weak lensing and its cross-correlations rely on high fidelity imaging from long baselines. With only AA* baselines, a high enough number density of resolved galaxies cannot be detected to measure the statistical signal.
Other observables will still be accessible for the AA* configuration, such as the measurement of the structure growth parameters or the baryonic feedback parameters.

Having available wide-area scanning and techniques such as on-the-fly,
much of the cosmology with \hi IM and continuum can be done using AA*. 
Despite the lower levels of precision on cosmological parameters due to the lower available sensitivity, AA* surveys are crucial to refine the data analysis pipelines and improve the control over possible systematics. 
With the full information obtained in AA4 configuration, these surveys will allow us to probe evolving dark energy, primordial non-Gaussianity and relativistic effects.

SKA surveys will have significant value in multiple critical ways. They will provide a new radio view of existing science probes that operate dedicated experiments in other wavebands. The success of these probes in answering fundamental science questions increasingly relies on independent verification from experiments with independent systematic uncertainties. SKA will provide unique information on the largest astronomical scales, on the evolution of the  baryon content of the Universe, and in redshift ranges that can otherwise not be measured. Finally, they will reach forward into discovery space for the truly unforeseen: to look for the solution to foundational questions of the history and behaviour of the Universe.

\bibliographystyle{abbrvnat-maxbibnames4}
\bibliography{chapter} % if your bibtex file is called example.bib

\end{document}